\documentclass[]{emulateapj}

\slugcomment{Submitted to The Astrophysical Journal}

\shorttitle{Galaxy Rotation and SMBH Binaries}
\shortauthors{K. Holley-Bockelmann \& F. M. Khan}

\begin{document}

\title{Galaxy Rotation and Rapid Supermassive Black Hole Binary Coalescence}

\author{Kelly Holley-Bockelmann\altaffilmark{1, 2}}
\author{Fazeel Mahmood Khan\altaffilmark{3}}

\affil{$^1$Vanderbilt University, Nashville, TN, USA, k.holley@vanderbilt.edu}
\affil{$^2$Fisk University, Nashville, TN, USA}
\affil{$^3$Institute of Space Technology (IST), Islamabad}

\begin{abstract}

During a galaxy merger, the supermassive black hole (SMBH) in each galaxy is thought to sink to the center of the potential and form a supermassive black hole binary; this 
binary can eject stars via 3-body scattering, bringing the SMBHs ever closer. In a static
spherical galaxy model, the binary stalls at a separation of about a parsec after ejecting all the stars in its loss cone -- this is the well-known {\it final parsec problem.}
However it has been shown that SMBH binaries in non-spherical galactic nuclei harden at a nearly constant rate until reaching the gravitational wave regime. Here we use a suite of direct $N$-body simulations to follow SMBH binary evolution in both corotating and counterrotating flattened galaxy models. For $N > 500K$, we find that the evolution of the SMBH binary is convergent, and is independent of the particle number. Rotation in general increases the hardening rate of SMBH binaries even more effectively than galaxy geometry alone. SMBH binary hardening rates are similar for co- and counterrotating galaxies. In the corotating case, the center of mass of SMBH binary settles into an orbit that is in a corotation resonance with the background rotating model, and the coalescence time is roughly few hundred Myr faster than a non-rotating flattened model. We find that  counterrotation drives SMBHs to coalesce on a nearly radial orbit promptly after forming a hard binary. We discuss the implications for gravitational wave astronomy, hypervelocity star production, and the effect on the structure of the host galaxy.

\end{abstract}


\section{Introduction}\label{sec-intro}

Supermassive black holes (SMBH)s, with masses between $10^{6}$ and $10^{10}$ solar masses, 
lie at the heart of nearly every galaxy~\citep[e.g.][]{kr95}. With a few notable exceptions, these SMBHs dwell within
stellar spheroids -- spiral bulges, ellipticals, and S0s -- and observationally the SMBH mass is highly correlated with properties of its host spherioid~\citep[e.g.][]{geb00, fer00, mcc13, gs15}. This seems to show that the evolution of the SMBH and its host are deeply tied, and innumerable observational and theoretical studies bear this out~\citep{she14, vol12, hop10, mic07, dim05}.

There is an emerging picture
that SMBH hosts all rotate to some degree.
In terms of spiral bulges, rotational  support is common. The early-type stars in our own galactic center have a net counterrotation of 120 km/sec~\citep{gen96,sch09}. 
The prototypical pseudobulges are rapidly-rotating, disky structures~\citep{Kormendy04}, and most classical bulges show some rotation as well~\citep{gad12}.

Low mass ellipticals and S0s are well-known to be disky and rotationally-supported~\citep[e.g.][]{Faber97, ben94, cap07}.   However, rotation may be pervasive in all early-type galaxies;  a volume-limited census of the nearest ~250 early-type galaxies ATLAS$^{\rm 3D}$ found that an astonishing 86 percent are fast rotators~\citep{ems11}, with less than 3 percent of the sample being described as non-rotating. One of these non-rotators was thought to be the giant cD galaxy M87~\citep{ems04, kra11}, and yet new IFU data reveal that even this canonical non-rotator has an unmistakable kinematically-distinct core~\citep{arn14, ems14}. In fact, nearly half of the ATLAS$^{\rm 3D}$  sample contains kinematically-decoupled cores~\citep{kra11} (see also~\citep{fra88, ben88, dav01, ems04} ), so it appears that SMBH hosts not only rotate, but that the rotational structure is quite complex.

It is easy, theoretically, to expect that rotation is practically ubiquitous in stellar spheroids.  Gas-rich formation scenarios give rise
to rotation, as do dry, non-radial, major mergers~\citep{ben92, kho09, boi11, tsa15}. Indeed, a non-rotating galaxy seems to be a special, and rare,
class that may be created exclusively through a slew of gas-poor minor mergers~\citep{naa14}.  
Since SMBH hosts clearly rotate, and since the link between the SMBH and its host is so well-established, any study of the dynamics with a SMBH host should consider rotation. This paper explores the evolution of binary SMBHs in a rotating, flattened stellar system, and its effect on the structure of the stellar host.

Binary SMBHs are expected to form within the galaxy core after a merger, and if the two 
SMBHs coalesce, they are arguably the most powerful sources of gravitational radiation
in the Universe~\citep{Hughes03}. One problem that plagues SMBHs is how they merge together. The binary can eject stars via 3-body scattering, bringing the SMBHs ever closer. Once the ejected stars extract enough energy from the binary orbit to shrink the separation to roughly milli-parsec scales, gravitational radiation dominates, and the SMBHs coalesce~\citep{BBR80}.  However, the problem is that analytical calculations and simulations of static, spherical galaxies show that the binary's orbital separation stalls before the SMBHs can plunge toward merger (see ~\citep{MM03} for a review). The root cause of this hang up is simply a lack of low-angular momentum stars capable of interacting with the binary via 3-body scattering. This theoretical bottleneck has become known as the infamous ``final parsec problem".  Recent work by several teams has shown that SMBHs can readily coalesce in more realistically-shaped galaxy models~\citep{ber06,kh11,Preto11,gm12,kh13}, because the stellar orbits in triaxial and flattened galaxies can continually replenish the SMBH binary loss cone~\citep[e.g.][]{Yu02, MP2004, khb06, va14, li14}.

SMBH binary dynamics has been studied in spherical rotating systems, but the resolution of the simulations (fewer than 100000 particles) could not accurately track the evolution of the binary to the gravitational wave regime. Experiments suggest that we need of the order of 1 million particles to resolve the true evolution of the binary beyond the hard binary stage. These previous studies primarily looked at the evolution of binary eccentricity and inclination with respect to the host~\citep{ses11, ama10, wan14}, and discovered that the eccentricity of the binary 
increased dramatically in counterrotating systems, while in corotating systems, the binary tends to circularize. 

Here we study SMBH binary dynamics and evolution in a flat rotating galaxy model with b/a = 0.8, which is quite a bit less flattened than a realistic galactic nucleus. For this model, we found an $N$-independent SMBH binary evolution in~\citet{kh13}. In this study we explore the dependence of SMBH binary evolution on particle number for various amounts of rotation of surrounding cusp. We also study the energy, angular momentum and eccentricity evolution of the SMBH binary and estimate the coalescence times by scaling our model to various observed nearby galaxies.

The paper is organized in the following way: In Section~2 we describe the initial conditions and numerical methods for our direct $N$-body simulations. The results for the evolution of the SMBH binary in rotating axisymmetric galaxy models are explained in Section~3. Finally, Section~4 concludes, discussing caveats and future work. 
  
\section{Initial conditions and numerical methods}\label{sec-model}

\subsection{The host galaxies and their SMBHs}

\begin{figure}

\centerline{
  \resizebox{0.95\hsize}{!}{\includegraphics[angle=0]{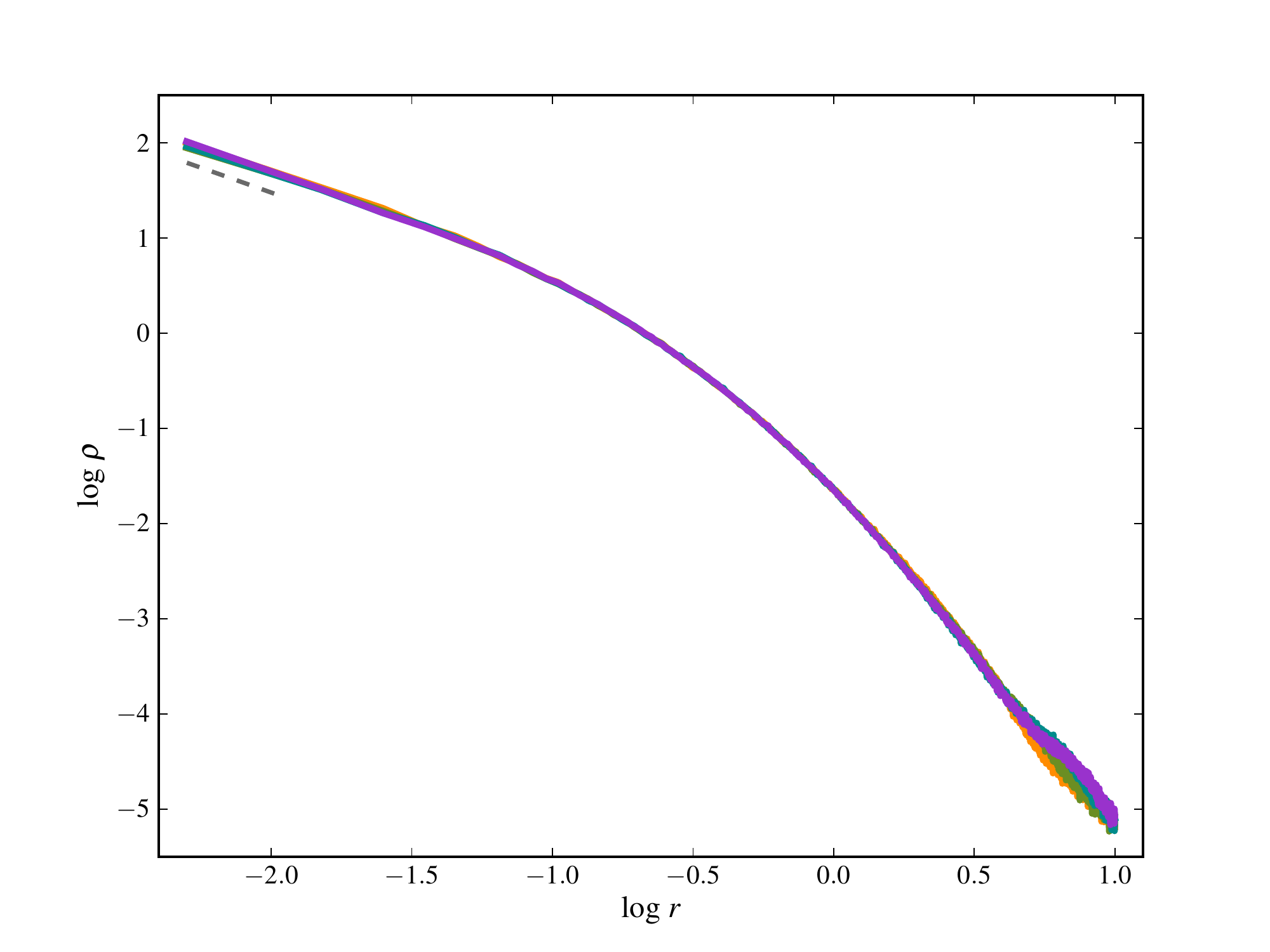}}
  }
\centerline{
  \resizebox{0.95\hsize}{!}{\includegraphics[angle=0]{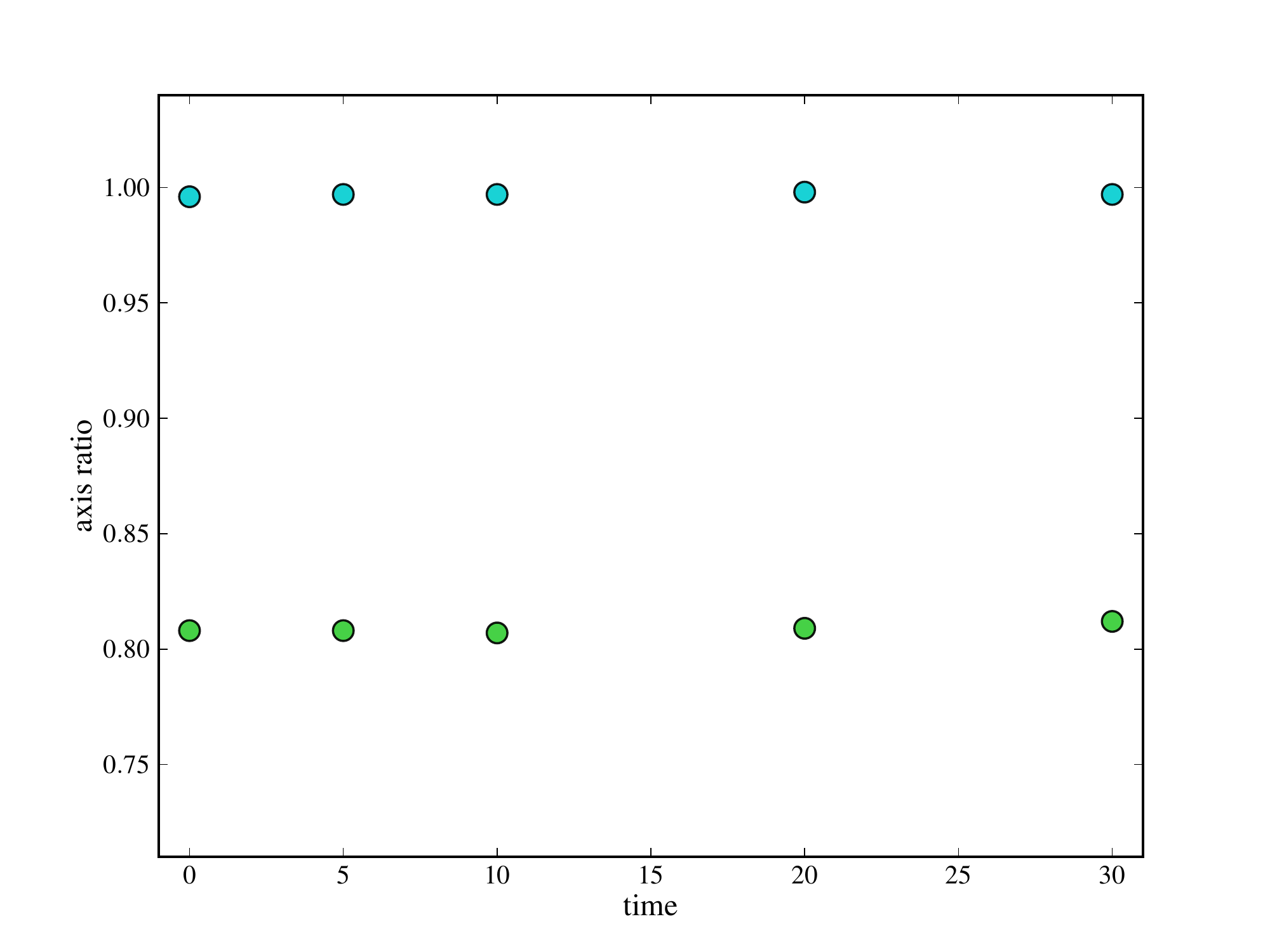}}
  }  
  \centerline{
  \resizebox{0.95\hsize}{!}{\includegraphics[angle=0]{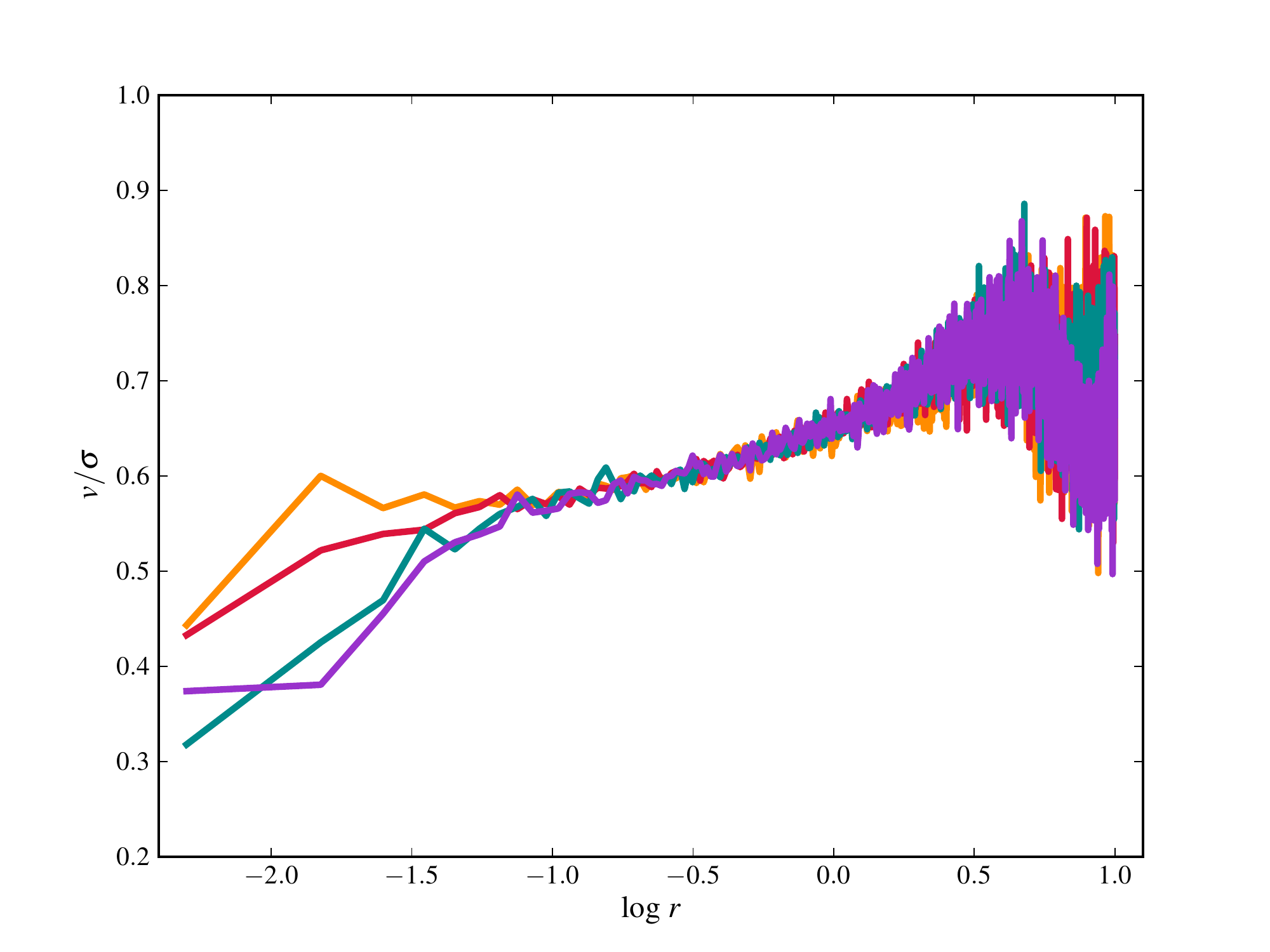}}
  }
\caption[]{Evolution of our flat rotating model (A0) in isolation.
Top Panel: Stellar density profile at various times. $\gamma = 1.0$ is the reference theoretical profile, shown as a dashed line. Middle Panel: Evolution of intermediate to major (b/a) and minor to major (c/a) axis ratio as measured at the half mass radius. Bottom Panel: Ratio of rotational velocity to the 3-d velocity dispersion as a function of distance at various times. In isolation, the rotational support in the center decreases over time as the central SMBH increases the velocity dispersion.
} \label{stab}
\end{figure} 

To isolate the effect of rotation, we used the Flat8 model in \citet{kh13} (hereafter KH13)  to generate our rotating models; this allows us to directly compare the SMBH binary orbital evolution results here to those in the identical non-rotating model. For this flattening ratio, we found that the SMBH binary evolved into the gravitational wave regime, and the evolution did not depend on particle number for $N>500K$.

We introduced rotation by flipping the z component of the angular momentum, $L_z$,  in the positive direction for a subset of particles which have negative $L_z$. Our fiducial model, the A series, flips every particle with negative $L_z$, while our B series only flips 50$\%$ of the negative $L_z$ particles, meaning that $75\%$ of the particles have a positive $L_z$.

Our galaxy models have an inner density slope, $\gamma$, of 1.0 and minor to major axis ratio of 0.8, measured at the half-mass radius. In model units, the total galaxy mass is 1 and the SMBH at the center of each model has a mass of 0.005. Because we adiabatically squeeze the model to generate its shape~\citep{khb01}, the system can and does change its shape with radius as it adiabatically adjusts to a live and changing SMBH-embedded potential. In addition, the scale radius slowly shrinks from 1.0 to 0.5; we could resize our model such that the scale radius is 1.0 again, but we instead choose to adopt physical units to reflect this increased central density. Discussion on scaling of our model to observed galaxies is given in section \ref{coal} and Table \ref{scale} scales the model to physical units.

To determine the stability of our rotating models, we ran them in isolation for 40 time units; this duration is about half the maximum evolution time of the SMBH binaries in this study. Figure \ref{stab} shows that both the density profile and half-mass axes ratio remain very stable for whole duration of the run. The kinematics are also fairly stable; inside the radius of influence, however, the system does become hotter, with $v/\sigma$ decreasing from ~0.6 to ~0.4 as the SMBH re-establishes its characteristic cusp in velocity dispersion. This decrease in rotational support affects only the innermost $\sim$ 5000 particles, but since it is also in the region of the model that exhibits a less flattened, but triaxial shape, it may well be that this region contains orbits that are less stable to rotation~\citep{dei11}. The orbit content of this model will be a subject for future study.

To explore the SMBH binary evolution in flat rotating galaxies, we introduce an equal mass secondary SMBH at a distance of 0.5 with 70 $\%$ of the galaxy's circular velocity at that initial separation. We investigated SMBH binary orbits that are corotating and counterrotating with the sense of the galaxy rotation.  Table \ref{TableA} describes parameters of our SMBH binary study. Note that we also evolved the SMBH binary in non-rotating spherical and flattened galaxy models with the same density profile to facilitate the comparison.

\begin{table}
\caption{Rotating Axisymmetric SMBH Binary Parameters} 
\centering
\begin{tabular}{c c c c c }
\hline
Run & $N$ & $\gamma$ & $c/a$ & $Rotation$\\
\hline
S0 & 1500k &	$1.0$& $1.0$& none\\
S1 & 1000k &	$1.0$& $1.0$& none\\
S2 & 500k &	$1.0$& $1.0$& none\\
A0 & 1500k &	$1.0$& $0.8$& corotating\\
A1 & 1000k &	$1.0$& $0.8$& corotating\\
A2 & 800k &	$1.0$& $0.8$& corotating\\
A3 & 500k & $1.0$& $0.8$& corotating\\
A4 & 250k & $1.0$& $0.8$& corotating\\[0.2ex]
B0 & 1500k & $1.0$& $0.8$& 0.75\% corotating\\
B1 & 1000k & $1.0$& $0.8$& 0.75\% corotating\\
B2 & 800k & $1.0$& $0.8$& 0.75\% corotating\\
B3 & 500k & $1.0$& $0.8$& 0.75\% corotating\\
B4 & 250k & $1.0$& $0.8$& 0.75\% corotating\\[0.2ex]
C0 & 1500k & $1.0$& $0.8$& counterrotating\\
C1 & 1000k & $1.0$& $0.8$& counterrotating\\
C2 & 800k & $1.0$& $0.8$& counterrotating\\
C3 & 500k & $1.0$& $0.8$& counterrotating\\

\hline
\end{tabular}\label{TableA}
\tablecomments{Column 1: Galaxy model. Column 2: Number of particles. Column 3: Central density slope $\gamma$.
 Column 4: Axes ratio. Column 5: Sense of galaxy rotation with respect to the SMBH binary initial orbit.}
\end{table}

The numerical methods and hardware used for this work is described in section 2.2 of KH13.

\section{SMBH BINARY EVOLUTION IN FLAT ROTATING GALAXY MODELS}\label{Resutls}

\begin{figure}

\centerline{
  \resizebox{0.95\hsize}{!}{\includegraphics[angle=0]{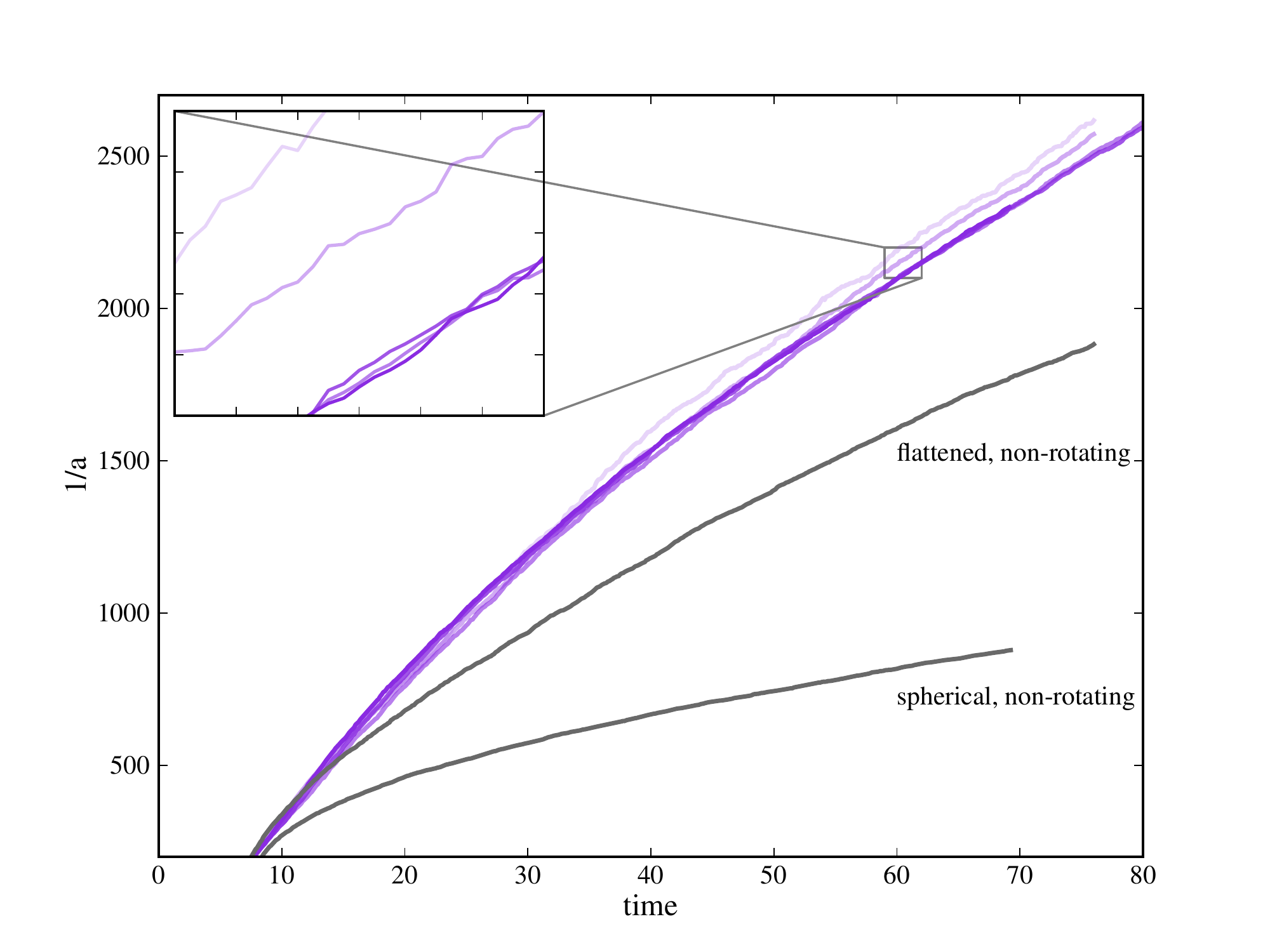}}
  }
\centerline{
  \resizebox{0.95\hsize}{!}{\includegraphics[angle=0]{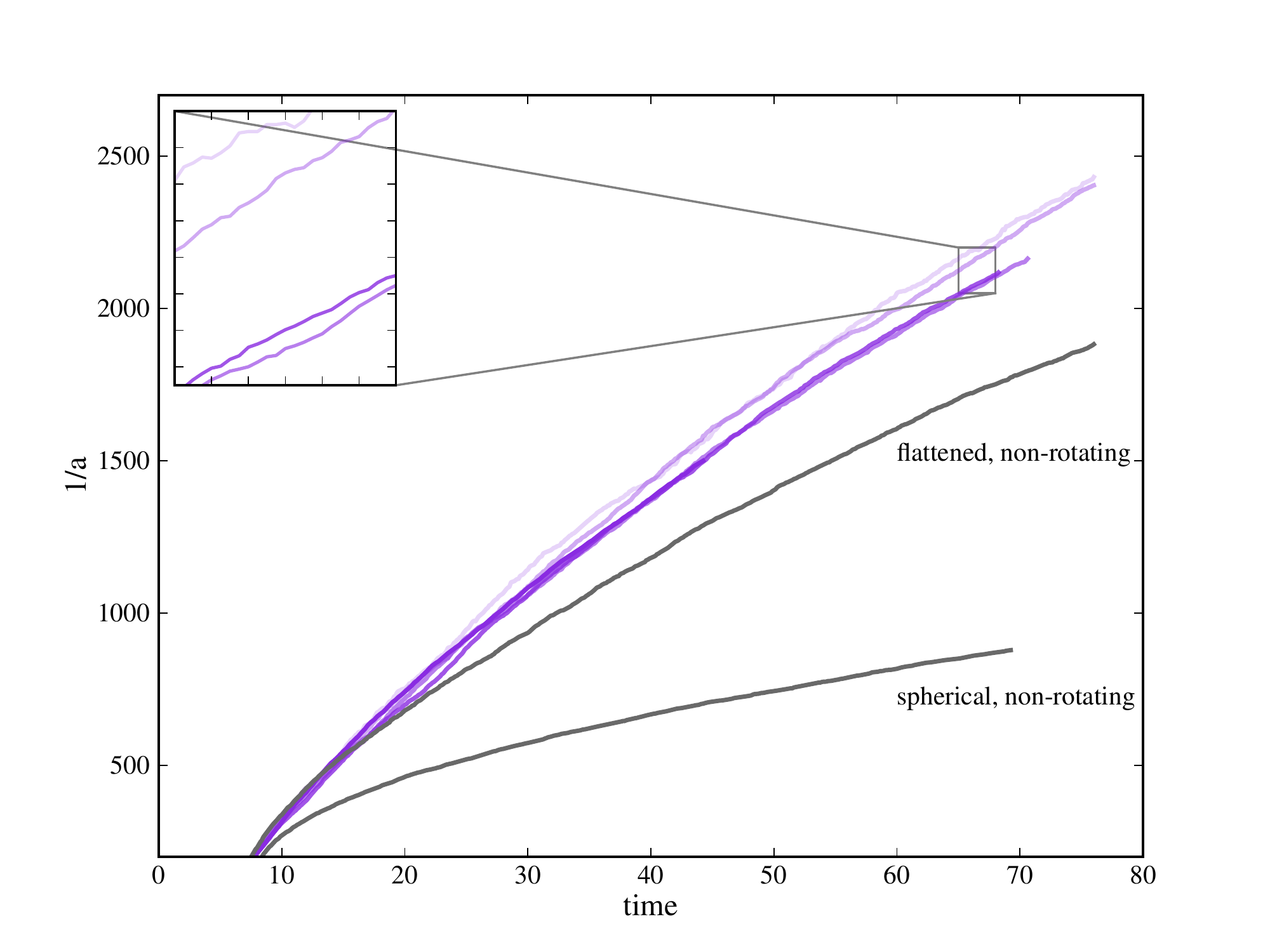}}
  }  
  \centerline{
  \resizebox{0.95\hsize}{!}{\includegraphics[angle=0]{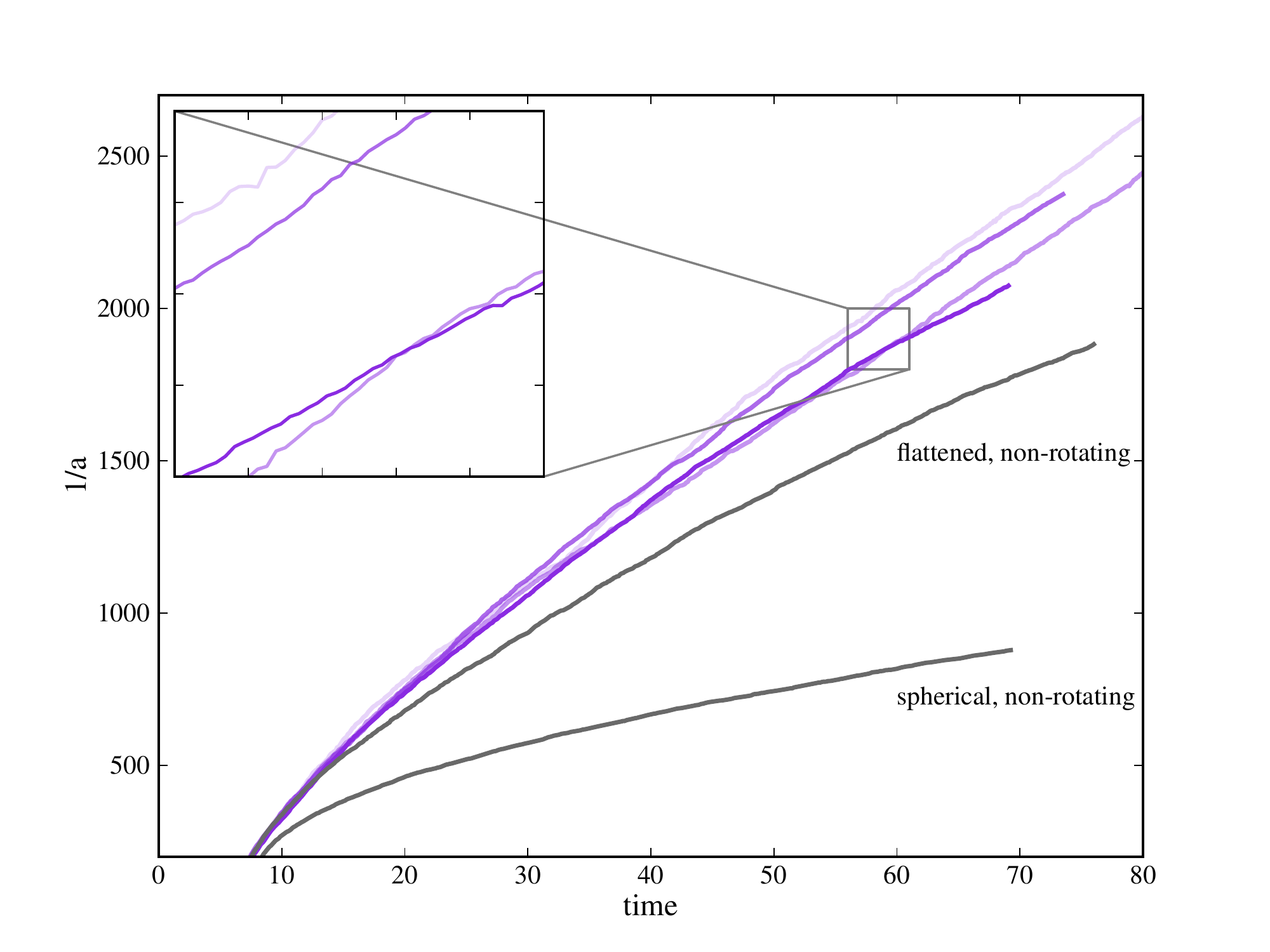}}
  }
\caption[]{Evolution of the semimajor axis of the SMBH binary for the A (top) B (middle) and C (bottom) model suites (see table \ref{TableA}). For comparison,  the grey lines represent non-rotating flattened  (uppermost grey) and non-rotating spherical (bottommost grey) one million particle models.  The opacity of the purple lines scale with the particle number -- the most transparent line has the fewest particles in the suite. 
} \label{rot1}
\end{figure}

Here we discuss the results of our numerical studies of SMBH binary evolution in flat rotating galaxy models. The top panel of Figure \ref{rot1} shows the evolution of inverse semi-major axis for the A models. We see that for N greater than 500K, the inverse semi major axis evolution is independent of N, unlike in \citet{va14}. To be conservative, we approach $N$ as high as 1.5 million, never used before in such a study. For reference, we also plot the 1/a evolution of a flat non-rotating model with 1 million from our previous study (KH13), as well as a 1.5 million particle run in spherical galaxy (S0) model with the same density profile as our rotating galaxy model. We can see that in rotating flat models, the SMBH binary evolves at a rate considerably faster than in mere flat galaxy models. We see $N$-independent evolution of the SMBH binary in flat rotating galaxy models for $N$ as large as 1.5 million. This points to a potential stellar dynamical solution to the final parsec problem within flattened galaxy models.

We also studied SMBH binary evolution in a galaxy with less dramatic bulk rotation; in this case, only 75 $\%$ particles corotate with the massive binary (models B in table \ref{TableA}). The middle panel of figure \ref{rot1} again shows that the model experiences $N$-independent evolution of 1/a. for $N \geq 800K$. However the binary coalesces at a slower pace when compared to models A for the same particle number.  For example, the convergent A models pass 1/a$=1500$ at 40 time units, while the
convergent B models take until nearly 50 time units to pass this same point.

The bottom panel of figure \ref{rot1} shows the evolution in the semi-major axis for the SMBH binary in a counterrotating orbital orientation (models C). Here, the binary orbit shrinks at slightly different rates than the previous models at various time intervals, though not with a clear trend. Af around 50 time units, the slope of 1/a line seems very similar for $N>500K$, and again we notice rapid evolution of 1/a when compared to flat and spherical models. However we have doubts that we are capturing the SMBH binary evolution accurately, because there is a clear dependence on particle number at later stages. In section \ref{coal} we discuss why we believe that SMBH binary coalescence is achieved in our counterrotating models, despite the lack of convergence in the model suite.

\begin{figure}

\centerline{
  \resizebox{0.95\hsize}{!}{\includegraphics[angle=0]{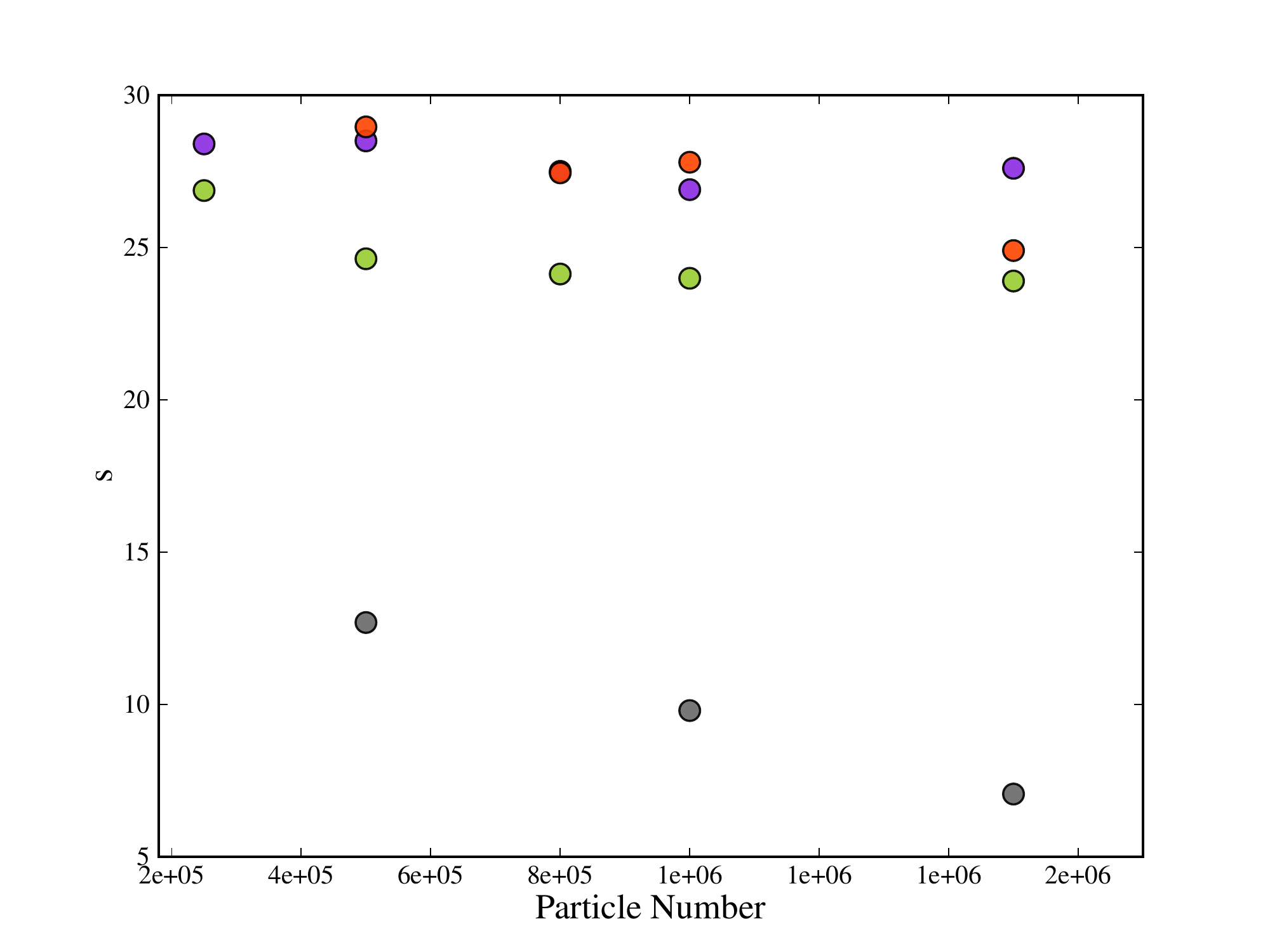}}
  }
\caption[]{
Hardening rates for all our numerical experiments. Grey points are for the spherical models (S); Purple represents models A; Green is for models B, and  the red points are for models C.
} \label{hard}
\end{figure}

For consistency between runs, we calculate the hardening rates $s$ for all our runs by fitting a straight line to the inverse semi major axis $s = \frac{d}{dt}(\frac{1}{a})$ in the interval 50-70 time units (fig \ref{hard}). Both corotating and counterrotating binaries (runs A \& C) have hardening rates of about $s \sim 28$ which is about $30 \%$ higher than mere flat models.  
The B runs have slightly lower values of $s$ $\sim$ 25 when compared to (runs A \& C). Overall, we find that $s$ in both co- and counterrotating models is approximately 4 times higher than in spherical models with exactly the same density profile, for our best resolved runs with 1.5 million particles.

\begin{figure}

\centerline{
  \resizebox{0.95\hsize}{!}{\includegraphics[angle=0]{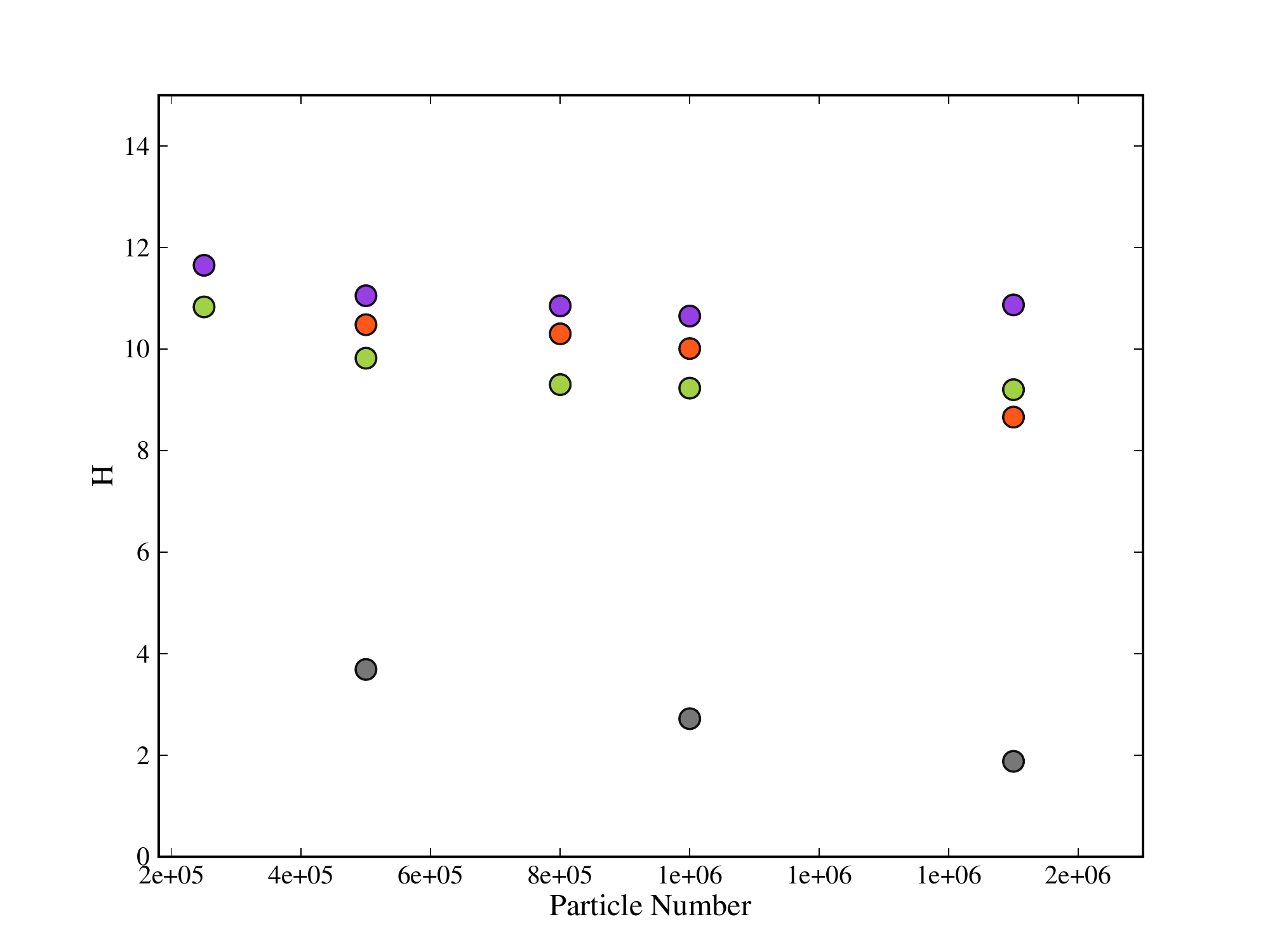}}
  }
\caption[]{
Hardening parameter $H$ for all our numerical experiments. Colors are as in figure~\ref{hard}.
} \label{h}
\end{figure}

For a spherical, homogeneous, isothermal background, the hardening parameter $H$ is related to hardening rate $s$ through $H = {s\sigma}/{G\rho}$  -- for the full loss cone regime, $H$ $\approx$ $15$~\citep{Q96,ses06}. We calculate $H$ by substituting values of $\rho$ and $\sigma$ at the influence radius, defined as sphere around the SMBH binary containing twice the mass of the binary in stars. Figure \ref{h} shows that for runs A2, A1 \& A0 with $N \geq 800k$, where the evolution is independent of N, the value of $H$ remains constant around $11$. The value of H from these $N$-body simulations is within 70\% of scattering experiments. For models B, again we see a constant value of $H$ $\approx$ $10$ for runs with $N \geq 800k$.  For spherical run S0 with greatest particle number $N$, $H$ $\simeq$ $1.9$ almost $8$ times smaller than what is predicted for a full loss cone in scattering experiments. We would like to point out that in our models, the background profile is not at all isothermal and there is also some ambiguity for where one should measure $\sigma$ and $\rho$ to make a fair comparison with scattering experiments. With this in mind, it is very encouraging that $H$ obtained from our study is well within a factor of two of idealized scattering experiments. 

\begin{figure}

\centerline{
  \resizebox{0.95\hsize}{!}{\includegraphics[angle=0]{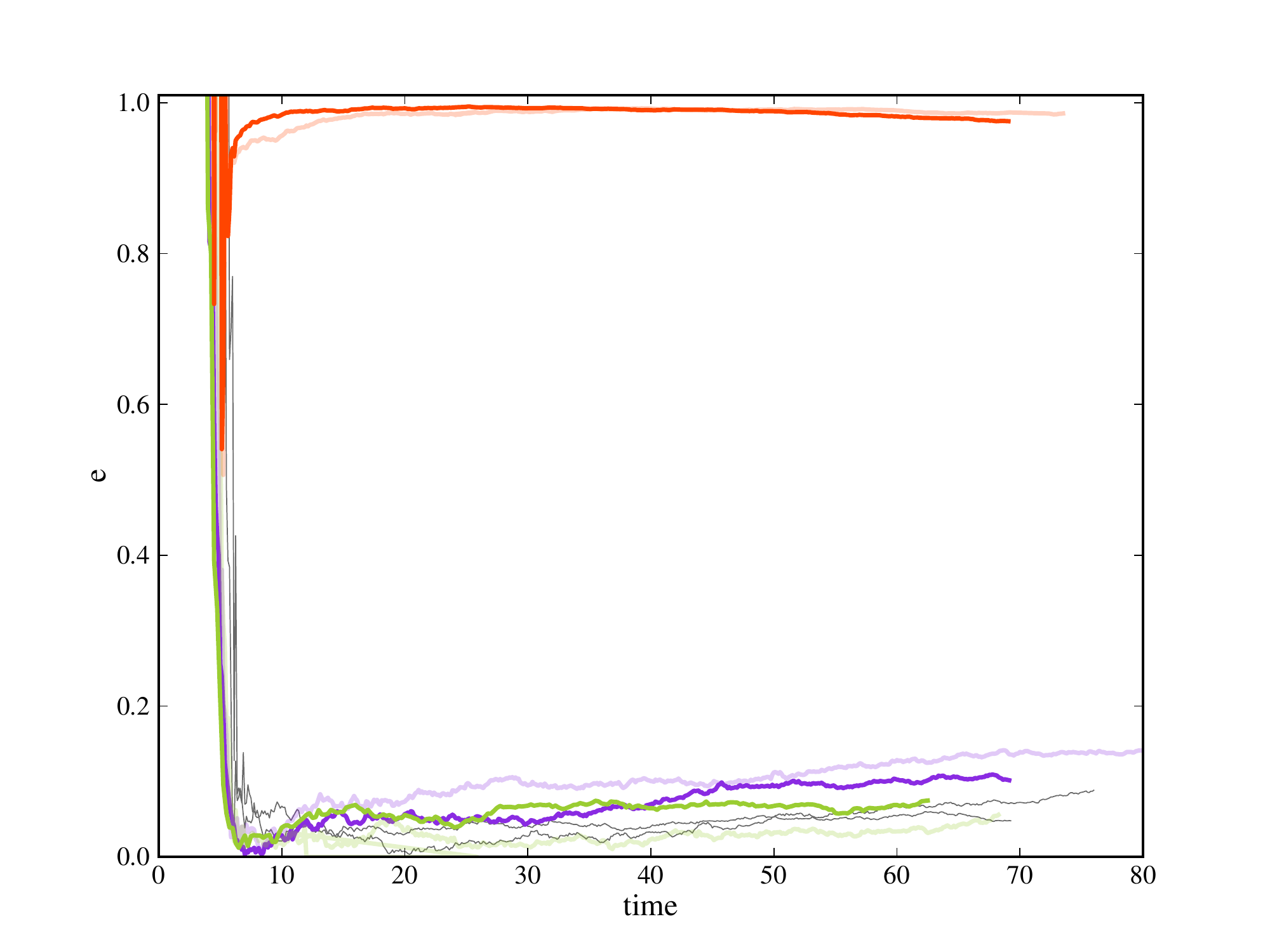}}
  }
\caption[]{
SMBH binaries eccentricity for models with $N \geq 1$ million. SMBH binaries in co-rotating, flat and spherical models have very small values of eccentricity $\sim$ 0.1 whereas  eccentricity approaches unity in counter-rotating models.  Here, the transparency 
represents the particle number; opaque lines are the $N=1.5$ million runs, while the fainter lines show results for $N=1$ million
particles. Orange lines show the counterrotating model; Green lines represent models B; Purple lines show models A, and the grey
lines are for the spherical and non-rotating axisymmetric models.
} \label{ec}
\end{figure}

Figure \ref{ec} shows the eccentricity evolution of SMBH binaries for all runs with $N \geq 1$ million. For corotating SMBH binaries, the eccentricities are consistently small ($e \sim 0.2$). For counter-rotating SMBH binaries, the eccentricity approaches $e \sim 1$ as soon as the binary forms. This is consistent with the findings of \citet{ses11}.

\begin{figure}

\centerline{
  \resizebox{0.95\hsize}{!}{\includegraphics[angle=0]{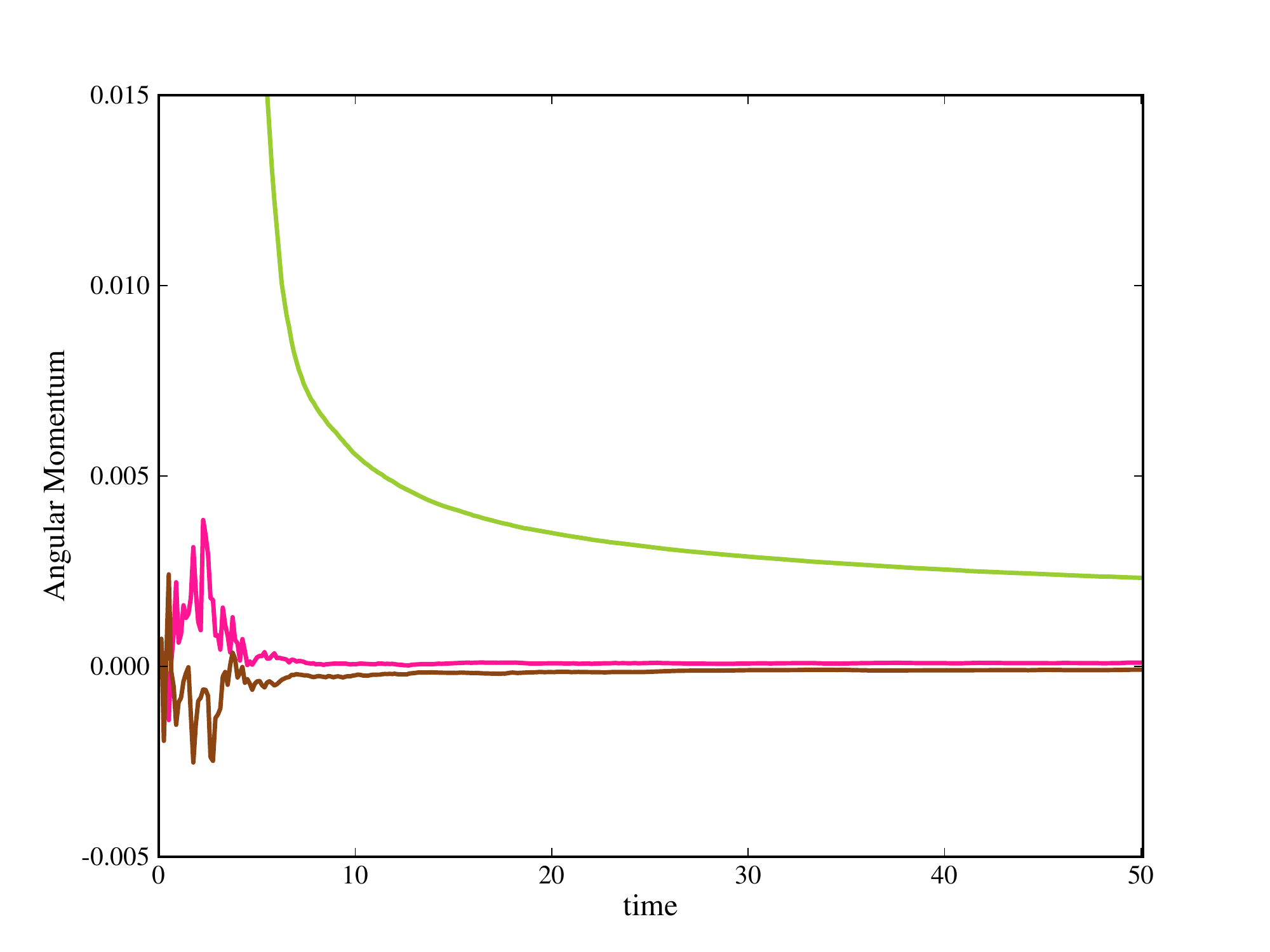}}
  }
\centerline{
  \resizebox{0.95\hsize}{!}{\includegraphics[angle=0]{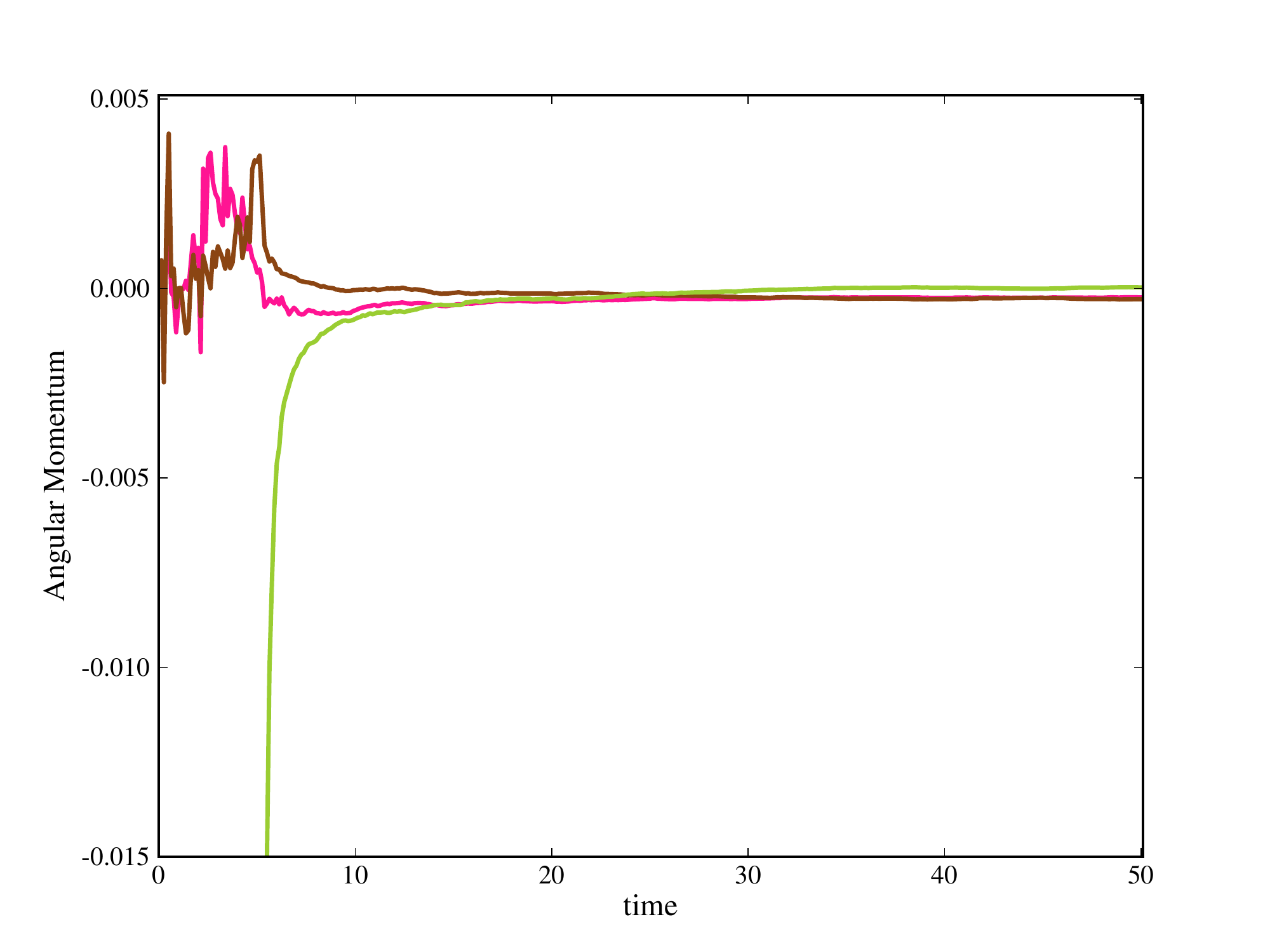}}
  }  
\caption[]{
Angular momentum evolution for corotating (top) and counterrotating (bottom) SMBH binaries. The green line is Lz, while the pink and brown lines are Lx and Ly, respectively.
} \label{ang1}
\end{figure}

The dichotomy in eccentricity behavior is borne out in a difference in the evolution of the angular momentum loss. Figure \ref{ang1} shows the angular momentum evolution of SMBH binaries for our best resolved co- and counterrotating models. 

It is clear that in the counterrotating case, the angular momentum loss is much more rapid. As we see from figure \ref{rot1}, the inverse semi-major axis evolution (and hence energy loss) is very similar for both co- and counterrotating models. This faster loss of angular momentum translates into a rapid rise in eccentricity.

\begin{figure}

\centerline{
  \resizebox{0.95\hsize}{!}{\includegraphics[angle=0]{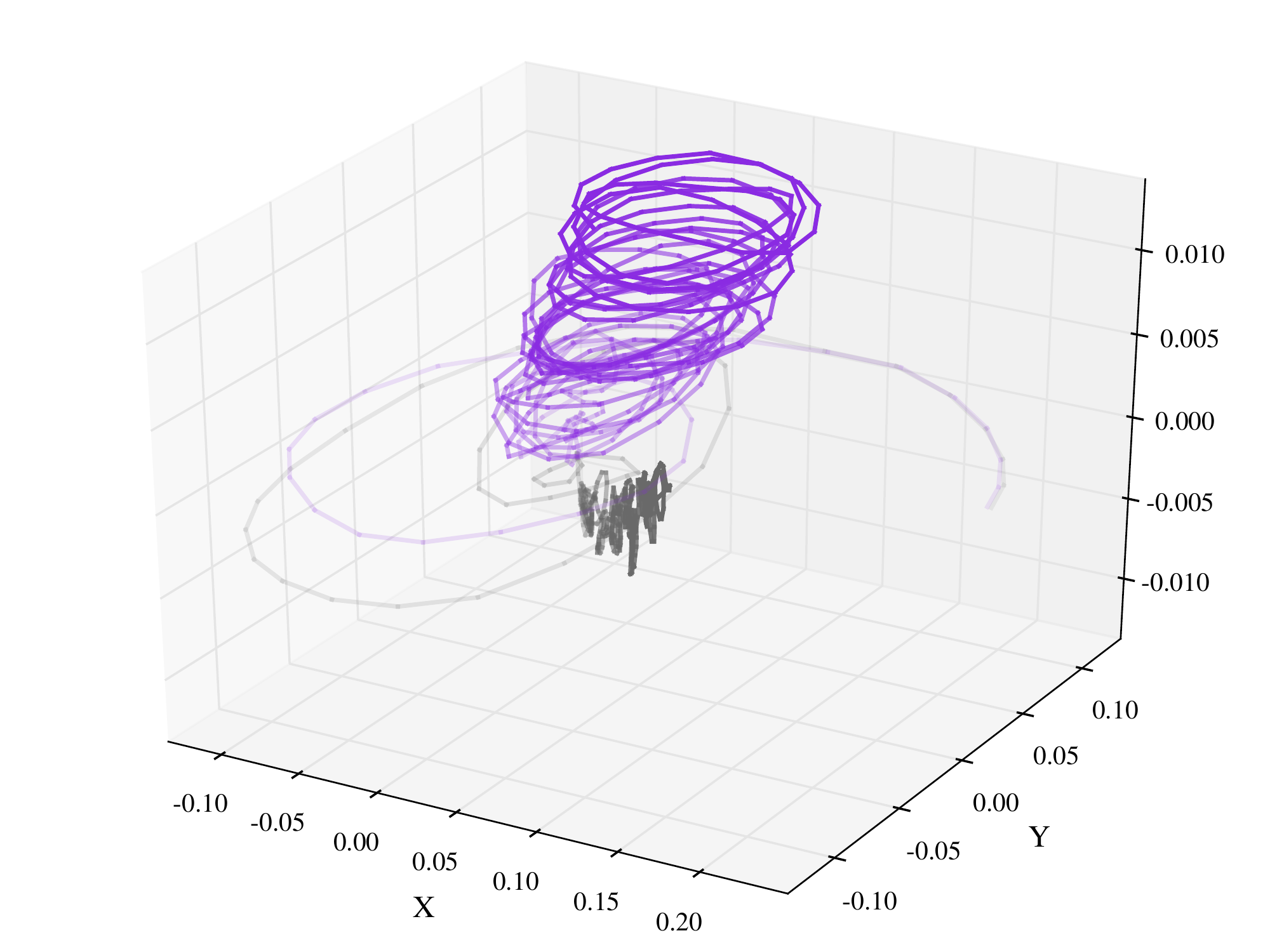}}
  }
\caption[]{
Motion of the center of mass of the SMBH binary in model A0 (purple) and the non-rotating flattened model (grey). Fainter colors indicate earlier epochs. Note that for model A0, the binary center of mass
settles into a roughly circular orbit about the galactic center with a radius of $\sim 0.05$ in model units, roughly the SMBH radius of influence. In contrast, the center of mass of the SMBH binary in the non-rotating model undergoes simple Brownian motion about the galactic center.
} \label{corot}
\end{figure}

We also investigated the center of mass motion of the binary in models A and C. Figure \ref{corot} shows the position of the center of mass throughout the run.

The trajectory of the SMBH binary center of mass is strongly effected by rotation. For non-rotating flattened models, the binary center of mass exhibits a small random walk characteristic of Brownian motion~\citep{cha02}. On the other hand,  the center of mass in the corotating system settles into a corotation resonance at the radius of influence, following a roughly circular orbit of radius $R_{\rm infl}$ nearly in the x-y plane.

\begin{figure}

\centerline{
  \resizebox{0.95\hsize}{!}{\includegraphics[angle=0]{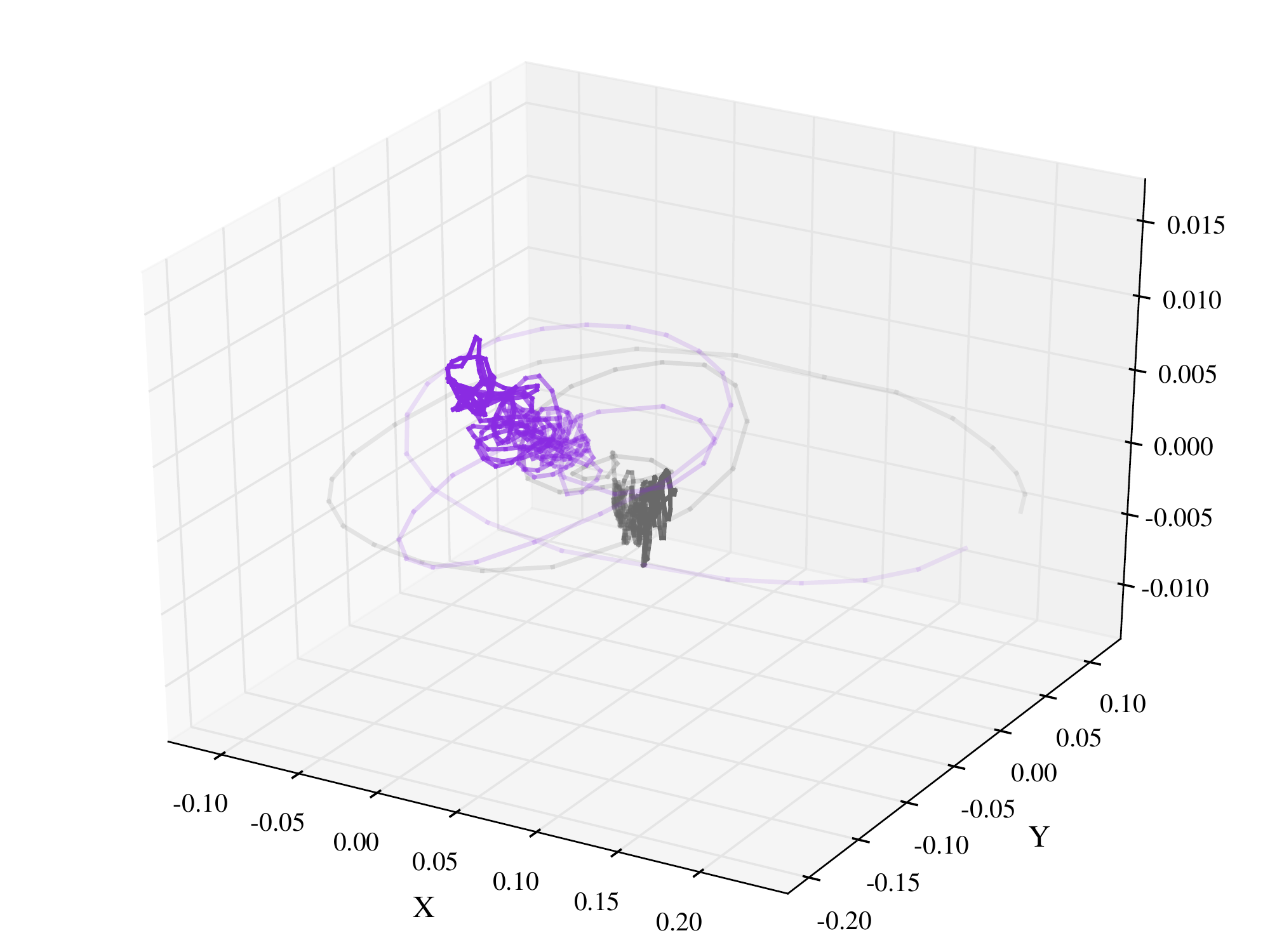}}
  }

\caption[]{
Motion of the SMBH binary center of mass in model C0 (purple) and the non-rotating flattened model (grey). The color transparency is as in figure~\ref{corot}. In our counterrotating model, the binary center of mass executes a random walk like the non-rotating model.
} \label{count-rot}
\end{figure} 

The counterrotating case, on the other hand, shows no binary orbital coupling, and the SMBH binary center of mass undergoes a random walk very similar to the non-rotating case.

\section{SMBH Binary Coalescence} \label{coal}

We also estimate the SMBH binary evolution for each case after the end of our simulation. We choose three Virgo cluster galaxies as reference to physically scale our models: M87, NGC4472, NGC4486A. In each case, the mass scale is set by the observed SMBH mass. For length scale,  we set the influence radius of the SMBH binary-embedded galaxy model to the size of the observed influence radius of reference galaxy model. NGC4472 may best represent our model density profile with its moderate central cusp, while the central core in M87 and the steep cusp of NGC4486A span the range of typical density cusps. Table \ref{scale} shows useful quantities for the physical scales in our models.

\begin{table}
\caption{Physical Scaling of our Models} 
\centering
\begin{tabular}{c c c c c c c c }
\hline
Galaxy & $M_{\bullet} (M_\odot)$ & $r_\mathrm{h}$(pc) & $T (\mathrm{Myr})$ & $L(\mathrm{kpc})$ & $M (M_{\odot})$ \\
\hline
M87 & $3.6 \times 10^9 $& $460$& $3.92$& $3.68$& $7.2 \times 10^{11}$\\
NGC4472 & $5.94 \times 10^8 $& $130$& $1.45$& $1.04$& $1.2 \times 10^{11}$\\
N4486A & $1.3 \times 10^7 $& $31$& $1.14$& $0.25$& $2.6 \times 10^9$\\

\hline
\end{tabular}\label{scale}
\tablecomments{Columns from left to right; (1) Reference galaxy, (2) Observed SMBH mass, (3) Observed SMBH radius of influence, (4) time unit, (5) length unit, (6) mass unit}
\end{table}

\begin{figure}
\bigskip

\centerline{
  \resizebox{0.95\hsize}{!}{\includegraphics[angle=0]{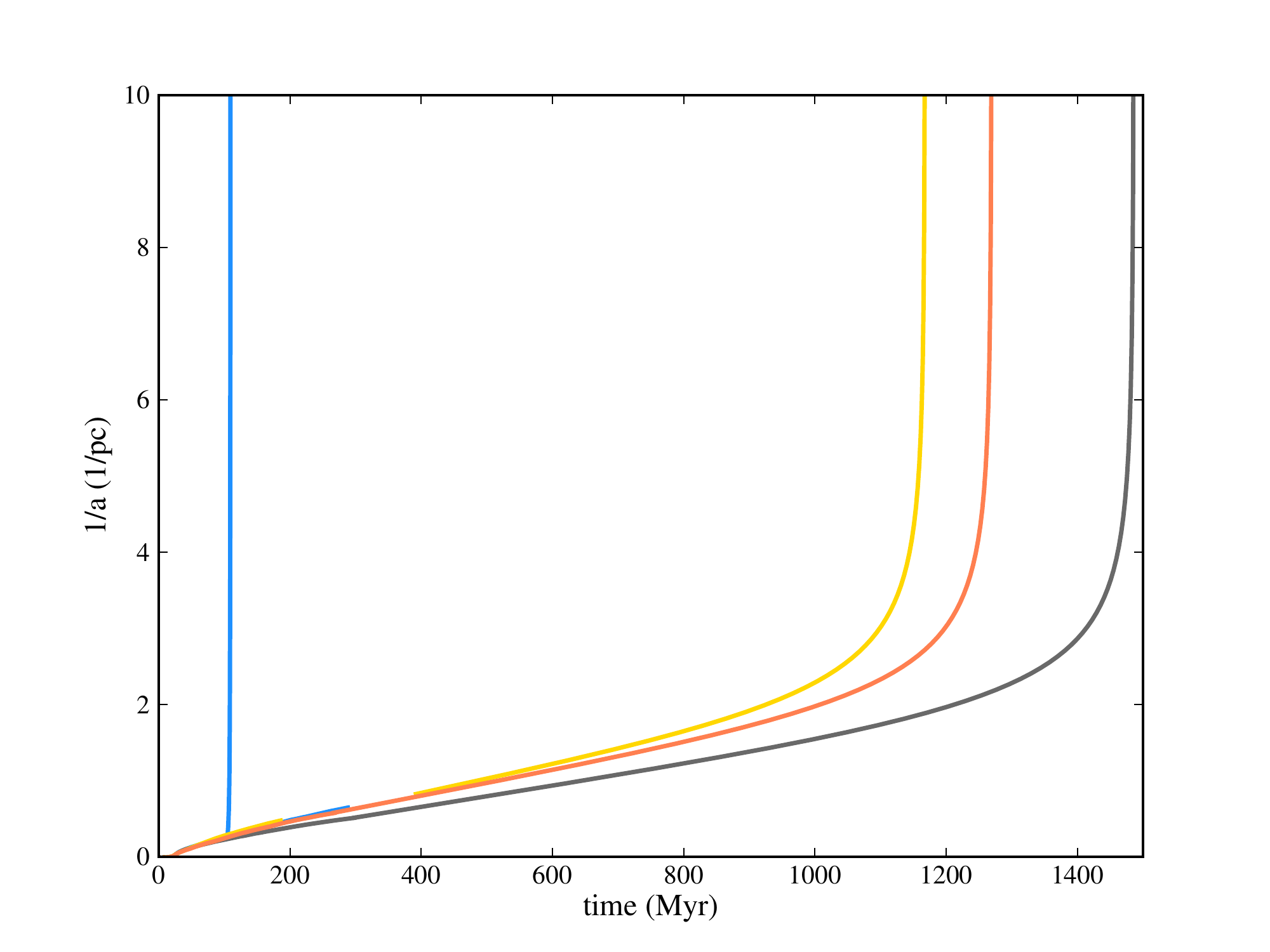}}
  }
\centerline{
  \resizebox{0.95\hsize}{!}{\includegraphics[angle=0]{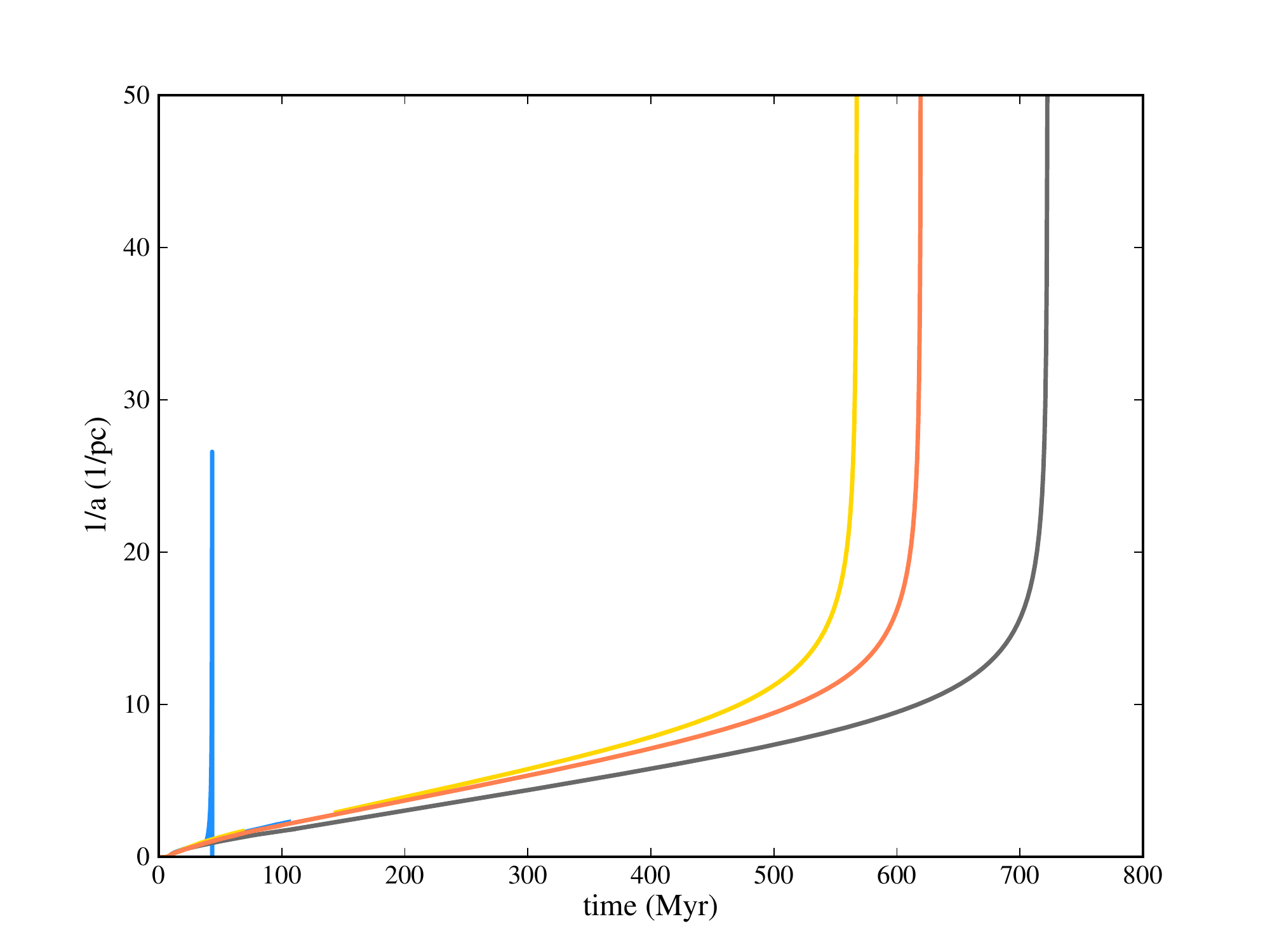}}
  }  
  \centerline{
  \resizebox{0.95\hsize}{!}{\includegraphics[angle=0]{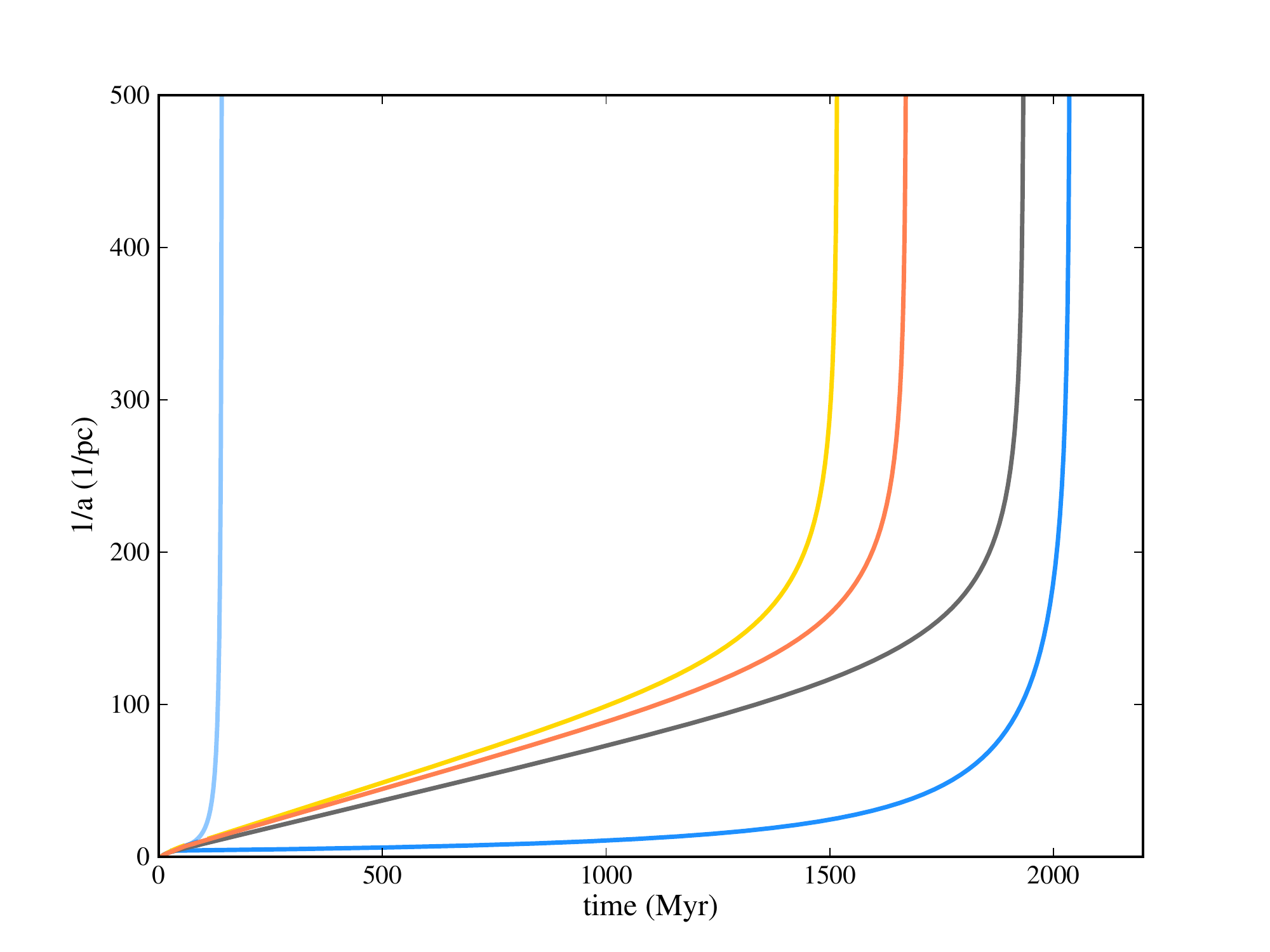}}
  } 
\caption[]{
Complete orbital evolution of the SMBH binary from formation to coalescence for runs A0 (yellow), B0 (orange), C0 (blue) and Flat (gray) when our galaxy model is scaled to M87 (top panel), NGC4472 (middle panel) and NGC4486A (bottom panel). In the bottom panel, the very different counterrotating timescales either include (short timescale with fainter hue) or exclude (longer timescale with stronger hue) stellar hardening.}
 \label{grav}
\end{figure}

Our technique for extrapolating the evolution of the SMBH binary beyond the endpoint of the simulation is explained in detail in section 4.3 of \citet{kh12b}. We choose the runs with highest particle number in each model (A0, B0, C0) for this extrapolation technique, and the evolution is shown in figure \ref{grav}. The top panel shows the SMBH binary evolution scaled to M87. 

In the non-rotating case, the SMBH binary coalesces in roughly 1.5 Gyr for this physical scaling. For corotating models A0 and B0, coalescence times are roughly 1.3 and 1.1 Gyr. For the counterrotating run C0, we evolve the SMBH binary from $1/a = 1000$ (see figure \ref{rot1}) at a system time T = 27, and we only consider hardening by gravitational waves; this is because we are not certain that the scattering results converge when the binary orbit is smaller than this. The SMBH binary coalesces in a mere 100 Myr -- essentially immediately -- due to its near radial eccentricity. The middle panel of figure \ref{grav} shows the SMBH binary evolution for NGC4472. Here, the SMBH binaries coalesce approximately two times faster than in M87; a case in point: the corotating model A0, the SMBH binary coalesces in roughly 500 Myr. Finally, the bottom panel of \ref{grav} scales to NGC4486A, and in this case the SMBHs merge in model A0 in about 1.5 Gyr while the SMBH binaries coalesce in almost 2 Gyr within the non-rotating model. Neglecting stellar hardening, the SMBH binary in the counterrotating case coalesces in roughly 2 Gyr.  However, if we assume that SMBH binary reaches an asymptotic hardening rate in C0, coalescence happens immediately after a hard binary forms. Clearly, SMBH binaries coalesce faster in rotating flattened models, but the mechanism behind the coalescence is very different depending on the sense of rotation. In corotating models, the rapid coalescence is due to higher hardening rates, but for counterrotating models high eccentricity drives the merger. Out of three representative galaxies, the SMBH binary coalescence time is shortest for NGC4472.

We show the coalescence time for different runs and physical scaling in Table \ref{time}. Technically, the clock starts here when two SMBHs form a pair with a separation $\simeq$ 10 influence radii (presumably after a major galaxy merger), and ends with the
coalescence of the binary from gravitational radiation.

\begin{table}
\caption{Physical Scaling of our Models} 
\centering
\begin{tabular}{c c c c c c c}
\hline
Run & $s$ & $H$ & $e$ & $T_{c,M87}$ & $T_{c,4472}$ & $T_{c,4486A}$\\
\hline
A0 & $27.7$& $10.87$& $0.17$& $1.17$& $0.57$& $1.52$\\
B0 & $23.90$& $9.20$& $0.06$& $1.27$& $0.62$& $1.67$\\
C0 & $23.6$& $8.66$& $0.99$& $0.11$& $0.04$& $2.04(0.14)$\\

\hline
\end{tabular}\label{time}
\tablecomments{Columns from left to right; (1) SMBH binary evolution run, (2) SMBH binary hardening rate, (3) Hardening parameter H, (4) SMBH binary eccentricity. (5) Coalescence time (in Gyr) when our model is scaled to M87, (6) NGC4472 (7) and NGC4486A.}
\end{table}	

\section{Structure of the merger remnant}

\begin{figure}

\centerline{
  \resizebox{0.95\hsize}{!}{\includegraphics[angle=0]{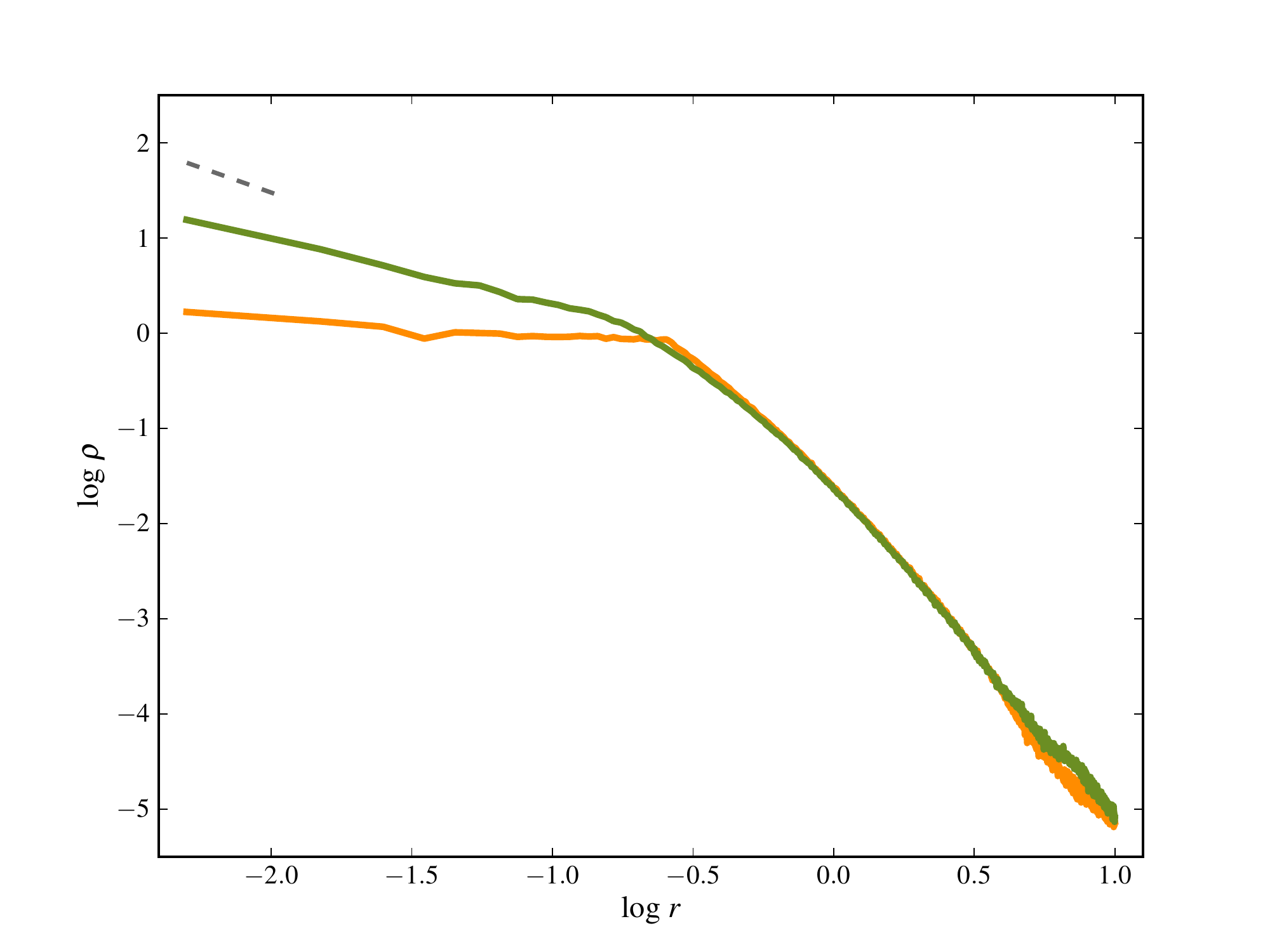}}
  }
\caption[]{
This is the initial (green) and final (orange) density profile of model A, showing the clear mass deficit
out to $r\sim0.3$ of 1.3 times the binary SMBH mass. 
} \label{scour}
\end{figure}

\begin{figure}
\centerline{
  \resizebox{0.95\hsize}{!}{\includegraphics[angle=0]{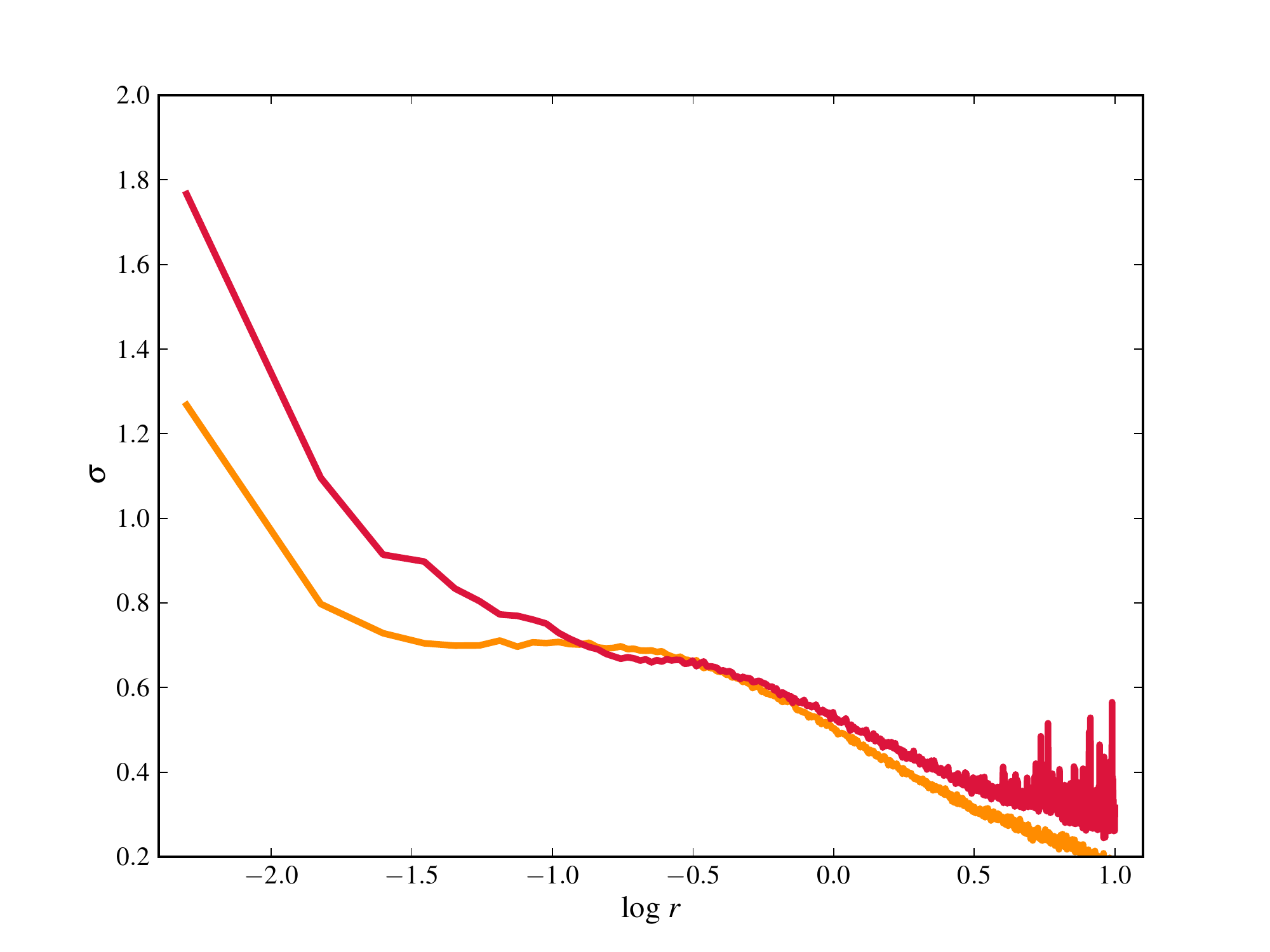}}
  }
\caption[]{
This is the initial (orange) and final (red) 3-dimensional velocity dispersion profile of model A, showing a sharp spike in the velocity dispersion at the end of the run. 
} 
\bigskip
\label{sigma}
\end{figure}

In this section, we go over the imprint of the binary black hole merger on the galaxy remnant.
As expected, figure \ref{scour} shows that the density cusp is scoured out by 3-body scattering 
as the black holes coalesce~\citep{gra04}. One puzzling result can be seen in figure \ref{sigma}, where the 
final velocity dispersion spikes; this is in contradiction to the dip in velocity dispersion expected
from the stellar hardening phase~\citep{Mei13}, and fully consistent with the velocity dispersion cusp 
expected in an equilibrium SMBH-embedded nucleus. Further study is needed, using simulations
with shorter snapshot output cadence, to help pinpoint the occurrence and longevity of this potential
kinematic signature of 3-body scattering.

\section{Conclusion}

We investigated the effect of bulk rotation on SMBH binary coalescence in N-body-generated flattened galaxy models. Overall, we found that rotation drives the SMBH binary more efficiently through the 3-body scattering phase, resulting in coalescence timescales
that are between 3 to 30 times faster than the same non-rotating model for co- and counterrotating models, respectively.
The 3-body scattering phase removes roughly 1.3 times the binary SMBH mass, scouring the density cusp out to about 1.1 kiloparsec if scaled to the M87 core.

We found that when the SMBH binary and the galaxy are corotating, the eccentricity remains low at approximately 0.1, while
counter-rotation acts to pump the SMBH binary eccentricity up to nearly one during the inspiral phase. Such a high eccentricity
enhances the coalescence of the SMBH binary, as is seen in many previous studies~\citep[e.g.][]{ses10,kh12a}. Though we caution that the
eccentricity behavior is not convergent even for 1.5 million particles, we suspect that the eccentricity will remain high in the convergent regime; when the SMBH is counterrotating, the abundance of retrograde orbits can extract angular momentum from the binary very efficiently, and secular dynamical anti-friction~\citep{mad12} torques the orbit so that it bleeds angular momentum. 
A systematic study is needed to gauge the degree of counterrotation and binary orbital plane alignment needed to pump the eccentricity into the nearly radial regime; if the binary eccentricity is very sensitive to minor degrees of counterrotation, then few mergers will linger in the 3-body scattering stage. For such high eccentricities, we should expect residual eccentricity to persist into the last few orbits in the gravitational wave regime; this will have profound implications for gravitational wave detection using waveform template matching.

\acknowledgments
We would like to thank Massimo Dotti for very helpful discussions during the early stages of this work. The simulations were conducted in part using the resources of the Advanced Computing Center for Research and Education at Vanderbilt University, Nashville, TN. 
KHB also acknowledges support from the NSF Career award AST-0847696.

\end{document}